# Enhancing *z* spin generation in trivial spin Hall materials for scalable, energy-efficient, field-free, complete spin-orbit torque switching applications

Qianbiao Liu[1], Lijun Zhu[1,2]*

[1]State Key Laboratory of Semiconductor Physics and Chip Technologies, Institute of Semiconductors, Chinese Academy of Sciences, Beijing 100083, China

[2]Center of Materials Science and Optoelectronics Engineering, University of Chinese Academy of Sciences, Beijing 100049, China

*ljzhu@semi.ac.cn

**Despite the remarkable efforts in the past two decades, it has remained a major challenge to achieve switching of perpendicularly magnetized spin-orbit torque devices in a scalable, energy-efficient, field-free, integration-friendly, and complete manner. Here, we report giant enhancement of *z* spin generation in low-resistivity spin Hall metal/FeCoB devices by alloying the spin Hall metal Pt with Ti and by electric asymmetry engineering. The dampinglike spin torques of *z* spins and *y* spins are enhanced by 6 and 3 times relative to that of conventional Pt/FeCoB and enable complete, record-low-power, deterministic switching of FeCoB devices with strong perpendicular magnetic anisotropy and high coercivity. The Pt$_{75}$Ti$_{25}$/FeCoB heterostructure also exhibits relatively low resistivity, wafer-scale uniform sputter-deposition on silicon oxide, good compatibility with magnetic tunnel junctions, and excellent thermal stability of exceeding 400 ºC. These results unambiguously establish the Pt$_{75}$Ti$_{25}$/FeCoB as the most compelling candidate for solving the bottleneck of scalable, energy-efficient, field-free, integration-friendly, and complete spin-orbit torque switching technologies. This work also provides a universal strategy for developing high-performance generators of *z* spin current and will stimulate the exploration of exotic spin currents by alloying "trivial" spin Hall materials.**

**Introduction**

Perpendicular magnetization is attractive for spin-orbit torque (SOT)-driven memory and computing with long data retention.[1] However, after the two-decade intensive efforts,[2-34] it has remained a challenge to switch a stable, uniform perpendicular magnetization by an in-plane current in a scalable, energy-efficient, magnetic-field-free, and integration-friendly manner. For example, a large charge current and a large in-plane magnetic field along the current direction (*x* direction) are typically required to switch a uniform perpendicular magnetization by the transverse spins (*y* spins, polarized in the *y* direction but flows in the *z* direction).[1,2] Spin current of perpendicular spins (*z* spins, both polarized and flows in the *z* direction)[22-34] has the potential to enable, by itself or in combination with *y* spin current, field-free switching of perpendicular magnetization. However, most discussions of the *z* spins have been limited to spin-rotatory interfaces of magnetic multilayers[23-25] that have low energy efficiency (see below) and crystallographically or magnetically low-symmetry single crystals[22,26-33] that can hardly be integrated into large-scale CMOS circuits. Importantly, most of the so-called "field-free" switching strategies can only yield incomplete switching of the magnetization and are incompatible with perpendicular FeCoB-MgO magnetic tunnel junctions, which questions the practical usefulness of the strategies.

Recently, the efficient generation of *z* spins has been demonstrated in Ta/Py bilayers that were sputter-deposited on an oxidized silicon wafer by tuning the spin Hall conductivity tensor ($\overleftrightarrow{\sigma_{SH}}$) of Ta via electric asymmetries[34] (see Fig. 1a for a simplified, schematical depiction of the transverse and perpendicular electric asymmetries). However, Pt devices only generate minimal $z$ spins[12] likely because the spin Hall conductivity tensor of pure Pt is robust against electric asymmetries[34]. Technologically, it is of particular importance to enhance $z$ spin generation in Pt since it already has the giant spin Hall conductivity for $y$ spins,[35] low resistivity, high spin-mixing conductance in contact with ferromagnets[36,37], and sputter-deposition on silicon oxide.[35] On the other hand, enhancing SOT exerted on the ferromagnet $Fe_{60}Co_{20}B_{20}$ is particularly interesting for its direct relevance to magnetic tunnel junctions in memory and computing technologies.

Here, we, for the first time, report giant enhancement of $z$ spin generation in sputter-deposited $Pt_{75}Ti_{25}/Fe_{60}Co_{20}B_{20}$ (PtTi/FeCoB) heterostructures by alloying Pt with Ti and by engineering the electric asymmetry via the layer thicknesses and geometry of the device. We show that the $z$ spins generated by the PtTi can enable highly efficient, deterministic, nearly full, record-low-power, external-magnetic-field-free switching of the FeCoB with strong perpendicular magnetic anisotropy (PMA) and high coercivity ($H_c$).

**RESULTS**

**Enhancing $z$ and $y$ spin generation by alloying**

Stimulated by the previous reports that the intrinsic spin Hall effect and thus the spin Hall ratio of Pt can be engineered significantly by alloying[35,37-41], we attempt to enhance the $z$ spin component of $\overleftrightarrow{\sigma_{SH}}$ by alloying Pt with another metal Ti. Magnetic bilayers of Pt 8/FeCoB 8 and PtTi 8/FeCoB 8 are fabricated on 4-inch silicon oxide substrates and patterned into spin-torque ferromagnetic resonance (ST-FMR) devices (see the optical microscopy image in Fig. 1b). As shown in Fig. 1c,d, the ST-FMR spectrum ($V_{mix}$) of each device is collected as a function of the in-plane field ($H$) while fixing the azimuth angle ($\varphi$) of $H$ relative to the rf current (the $x$ direction) asymmetrically injected into the device using the two-terminal geometry (Fig. 1b). The ST-FMR spectrum includes a symmetric ($S$) and anti-symmetric ($A$) components which can be separated from fits of the spectrum to the equation

$$V_{mix} = S \frac{\Delta H^2}{\Delta H^2+(H-H_r)^2} + A \frac{\Delta H(H-H_r)}{\Delta H^2+(H-H_r)^2}, \quad (1)$$

where $\Delta H$ is the resonance linewidth and $H_r$ the resonance field. In general, $S$ and $A$ vary with $\varphi$ as[34]:

$S=(S_{DL,y}+S_{art})\sin2\varphi\cos\varphi+S_{DL,x}\sin2\varphi\sin\varphi+S_{FL+Oe,z}\sin2\varphi$, (2)

$A = A_{FL+Oe,y} \sin2\varphi \cos\varphi + A_{FL,x} \sin2\varphi \sin\varphi + A_{DL,z} \sin2\varphi$. (3)

In Eq. (2), $S_{DL,y}$ is the contribution from the dampinglike SOT of $y$ spins from the spin Hall metal,[22,42] $S_{art}$ includes possible bulk SOT of self-induced $y$ spins,[43,44] spin pumping-inverse spin Hall voltage of the $y$ spins and the Nernst voltage induced by Joule heating,[45] $S_{DL,x}$ is the dampinglike SOT of the $x$ spins, and $S_{FL+Oe,z}$ is from the field-like torque of the $z$ spins and the perpendicular Oersted field.[10] In Eq. (3), $A_{FL+Oe,y}$ is from the field-like torque of $y$ spins and the transverse Oersted field, $A_{FL,x}$ is the field-like torque of the $x$ spins, and $A_{DL,z}$ is the dampinglike torque of the $z$ spins.[22] From the fits of the $\varphi$ dependences of $S$ and $A$ to Eq. (2) and Eq. (3) in Fig. 1e, the Pt 8/FeCoB 8 exhibits a minimal dampinglike torque of $z$ spins but a sizable $S_{FL+Oe,z}$ signal predominantly from the perpendicular Oersted field, which is consistent with previous experiments on Pt/Co and Pt/Py devices.[12] In contrast, the PtTi 8/FeCoB 8 device exhibits much greater $S_{FL+Oe,z}$ and $A_{DL,z}$ signals (Fig. 1f), indicating the presence of significantly enhanced dampinglike and fieldlike SOTs of $z$ spins in the PtTi devices. There is no $x$ spin signal ($S_{DL,x} \approx 0$) in the studied devices as is usually the case.[12,34]

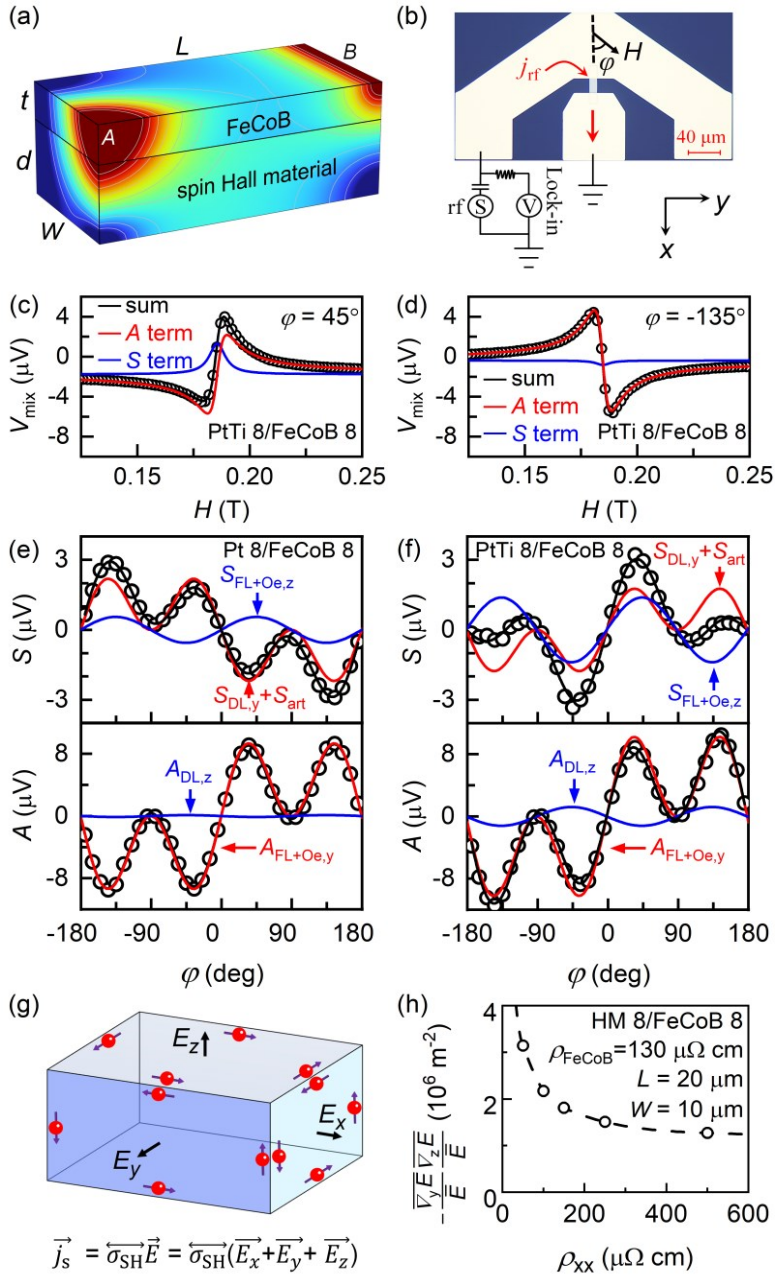

**Fig. 1 Enhancing z spin generation by alloying.** (a) Schematic depict of the transverse and perpendicular electric asymmetries in a spin Hall metal/FeCoB bilayer strip when a charge current is injected from sport A and flows out from side B. (b) Optical microscopy image of a ST-FMR device and the two-terminal measurement configuration. Representative ST-FMR spectra of the PtTi 8/FeCoB 8 device for the azimuth angle ($\varphi$) of the field at (c) $\varphi = 45°$ and (d) $\varphi = 45°$. Dependences on $\varphi$ of $S$ and $A$ for (e) the Pt 8/FeCoB 8 and (f) the PtTi 8/FeCoB 8 devices. (g) Spin current generation by the spin Hall conductivity tensor $\overleftrightarrow{\sigma_{SH}}$ and arbitrary electric field $\vec{E}$. (h) Finite-element simulation result of $-\frac{\overline{\nabla_y E}}{\overline{E}}\frac{\overline{\nabla_z E}}{\overline{E}}$ vs the resistivity of the heavy metal (HM). The blue curves represent the contributions of $z$ spins and perpendicular Oersted field, the red curves represent the contributions of $y$ spins and transverse Oersted field. In (c)-(f) and (h), $L = 10$ μm, $W = 20$ μm.

Following the method we discuss in the next section, $\xi_{DL,z}^{j}$ is estimated as 0.0016 for the Pt 8/FeCoB device and 0.0119 for the PtTi 8/FeCoB device ($W = 10$ μm, $L = 20$ μm). According to the drift-diffusion spin transport[35,46], the SOT exerted on a light ferromagnet by a $y$ or $z$ spin current from the spin Hall effect can be written as $\xi_{DL,y(z)}^{j} = T_{int}\sigma_{SH,y(z)}\rho_{xx}$, where $T_{int}$ is the spin transparency of the interface, $\sigma_{SH,y(z)}$ the spin Hall conductivity element of $y$ spins ($z$ spins), and $\rho_{xx}$ the resistivity of the spin Hall metal. Since $T_{int}$ is expected to be similar (≈0.50) for both the Pt 8 and PtTi 8 devices[37,47], we attribute the enhancement of $z$ spin generation by alloying Pt with Ti to a significant tuning of the spin Hall conductivity tensor of the Pt (i.e., the enhancement of $\sigma_{SH,z}$) and the enhancement of the resistivity ($\rho_{xx}$ is ≈ 100 μΩ cm for the 8 nm PtTi and 19 μΩ cm for the 8 nm Pt). We stress that $\overleftrightarrow{\sigma_{SH}}$ must be altered by alloying because the 5 times resistivity enhancement cannot, by itself, explain the more than tenfold enhancement of $z$ spin torque.

We also point out that the enhancement of the $z$ spin generation cannot be attributed to any current tilting itself. Without the altering the spin Hall conductivity tensor, a tilted current flow cannot generate any $z$ spin current. As shown in Fig. 1g, the spin current $\vec{J_s}$ generated by an arbitrary electric field $\vec{E}$ can be written as $\vec{J_s} = \overleftrightarrow{\sigma_{SH}} \vec{E} = \overleftrightarrow{\sigma_{SH}} (\vec{E_x}+\vec{E_y}+\vec{E_z})$, while none of the three components of the electric field ($\vec{E_x}, \vec{E_y}, \vec{E_z}$) can generate a spin current both polarized and flowing in the $z$ direction. Moreover, our finite-element simulation in in Fig. 1h indicates that the electric field gradients (characterized as the product of relative transverse and perpendicular electric field gradients, $\frac{\overline{\nabla_y E}}{\overline{E}} \frac{\overline{\nabla_z E}}{\overline{E}}$, the simulation method is described in the Methods and ref. 34), thus the tilting of the current flow, decreases rather than increases upon increase of the resistivity of the spin Hall metal.

**Enhancing $z$ spin generation by electric asymmetry engineering**

We show below that the $z$ spin generation can be significantly enhanced by engineering the electric symmetries of the device (Fig. 1a) via the thickness of the spin Hall metal ($d$), the thickness of the magnetic layer ($t$), the width ($W$) and length ($L$) of the magnetic strip of the PtTi/FeCoB. The principle is simply that the emergency of the spin Hall conductivity of $z$ spins is directly relevant to the electric symmetries of the devices (exactly zero in the absence of electric symmetries). Figure 2a,b shows the experimental results of dampinglike torque efficiency of the $z$ spins ($\xi_{DL,z}^{j}$) and the finite-element simulation results of the degree of the electric asymmetries as a function of $d$, $t$, $W$, and $L$. As discussed previously[34], $\xi_{DL,z}^{j}$ can be estimated following

$$\xi_{DL,z}^{j} = \xi_{DL,y}^{j} A_{DL,z}/S_{DL,y}\sqrt{1 + M_{eff}/H_r}, \quad (4)$$

where $M_{eff}$ is the effective magnetization as determined from the rf frequency dependence of $H_r$ (Fig. S1). The dampinglike SOT efficiency of $y$ spins ($\xi_{DL,y}^{j}$) is 0.110 ± 0.001 for the PtTi 4/FeCoB $t$ ($\rho_{xx}$ ≈ 120 μΩ cm for 4 nm PtTi), 0.091 ± 0.002 for the PtTi 8/FeCoB $t$ ($\rho_{xx}$ ≈ 100 μΩ cm for 8 nm PtTi), 0.050 ± 0.001 for the Pt (4)/FeCoB and 0.030 ± 0.003 for the Pt 8/FeCoB $t$ ($\rho_{xx}$ ≈ 19 μΩ cm for 8 nm Pt), as determined from the inverse intercept of the linear fit[48] of $1/\xi_{FMR,y}$ vs $1/t_{FeCoB}$ (Fig. 2c),

$$\frac{1}{\xi_{FMR,y}} = \frac{1}{\xi_{DL,y}^{j}}(1 + \frac{\hbar \xi_{FL,y}^{j}}{e\mu_0 M_s t_{FM} d_{HM}}). \quad (5)$$

Here, $e$ is the elementary charge, $\hbar$ the reduced Planck's constant, $\mu_0$ the permeability of vacuum, $M_s$ the saturation magnetization, $\xi_{FL,y}^{j}$ the fieldlike torque efficiency of $y$ spins, and $\zeta_{FMR,y}$ is defined as[48]

$$\zeta_{FMR,y} \equiv \frac{S_{DL,y}}{A_{FL,y}} \frac{e\mu_0 M_s t_{FM} t_{HM}}{\hbar} \sqrt{1 + M_{eff}/H_r}. \quad (6)$$

As shown in Fig. 2a, $\xi_{DL,z}^j$ can be increased significantly via increasing the PtTi thickness, the FeCoB thickness, and the width, and by decreasing the length, which is qualitatively consistent with variations of the simulated geometry-enhanced electric asymmetries in Fig. 2b. We also note that the dependences of $S_{FL+Oe,z}/S_{DL,y}$ on $d$, $t$, $W$, and $L$ (Fig. S2) are quite similar to those of $\xi_{DL,z}^j$, suggesting an enhancement of field-like torque of the $z$ spins and perpendicular Oersted field. We attribute the observed enhancement of the z-spin torques by the device geometry to an enhancement of the z spin Hall conductivity ($\sigma_{SH,z}$) by improved electric asymmetries. The strong dependences of the $z$-spin torque on the layer thicknesses, lateral geometry, and contact of the device unambiguously reveal that the $z$ spin torque is not from any interface effects, which is different from the mechanism of a previous report.[33] The maximum value of $\xi_{DL,z}^j$ is 0.015 for the optimal PtTi 8/FeCoB 8 device with $W = 15$ μm and $L = 10$ μm. Further enhancement of $\xi_{DL,z}^j$ is possible if electric asymmetries could be enhanced by fine geometry optimization via, e.g., layer thicknesses and contact design.

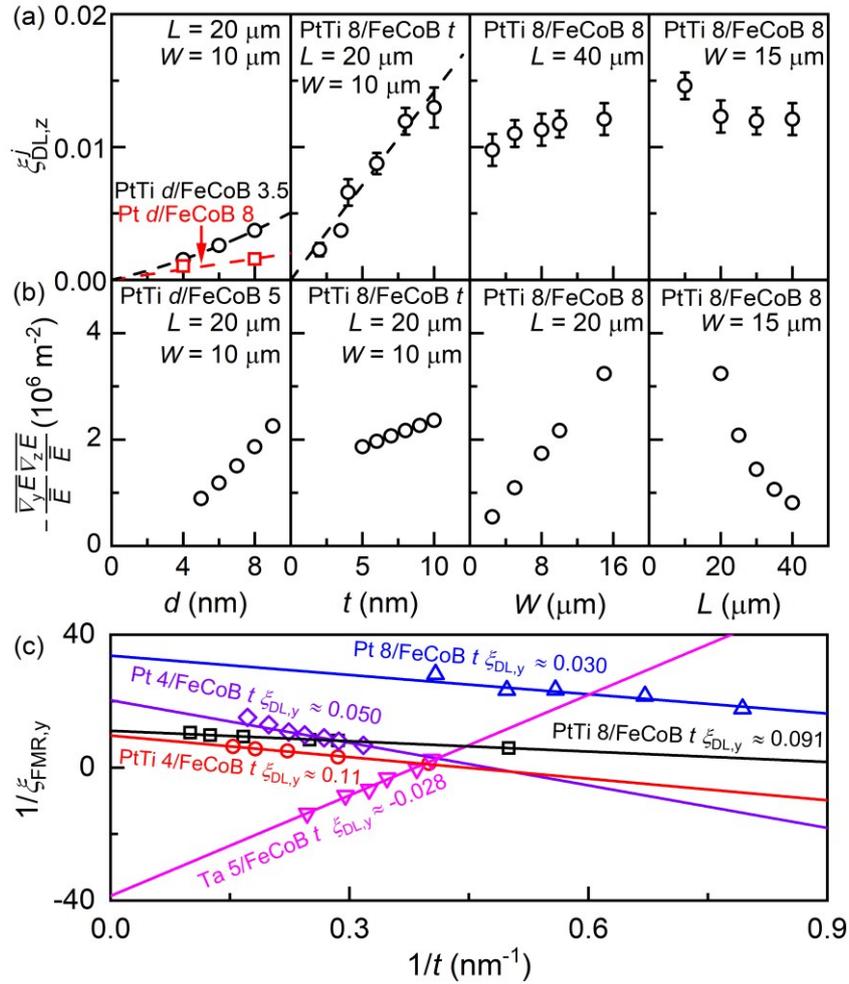

**Fig. 2 Spin-orbit torque efficiencies of $y$ spins and $z$ spins.** Dependences of (a) $\xi_{DL,z}^j$ and (b) $-\frac{\overline{\nabla_y E}}{\overline{E}} \frac{\overline{\nabla_z E}}{\overline{E}}$ of the PtTi $d$/FeCoB $t$ devices on the PtTi thickness $d$ ($t$ = 3.5 nm, $W$=10 μm, $L$=20 μm), and the FeCoB thicknesses $t$ for ($d$ = 8 nm, $W$=10 μm, $L$=20 μm), the width $W$ ($t = d = 8$ nm, $L$=20 μm), and the width $L$ ($t = d = 8$ nm, $W$=15 μm). The red squares plot $\xi_{DL,z}^j$ for the Pt 8/FeCoB. **(c)** $1/\xi_{FMR,y}$ vs $1/t$ for the PtTi 4/FeCoB $t$, PtTi 8/FeCoB $t$, Pt 4/FeCoB $t$, and Pt 8/FeCoB $t$ ($W$=10 μm, $L$=20 μm).

**Technological impacts**

Technologically attractive devices should simultaneously have good compatibility with large-scale integration with CMOS and magnetic tunnel junctions, relatively low resistivity, high SOT efficiencies of $z$ spins and $y$ spins, deterministic energy-efficient switching, and high coercivity at the same time. As shown in Fig. 3a, the PtTi-based z-spin device allows for wafer-scale uniform sputter-deposition on commercial Si/SiO$_2$ substrates, which is critical for large-scale integration into CMOS circuits and in sharp contrast to other field-free strategies requiring antiferromagnetic or piezoelectric single crystals[7,10,14-16,26-31] or the composition[3,17-19] or thickness wedges[3,21]. The PtTi-based z-spin devices with the magnetic FeCoB layer on the top also show great promise for integration with high-performance magnetic tunnel junctions.[1]

As summarized in Fig. 3b, the PtTi 8/FeCoB 8 has much greater $\xi^j_{DL,z}$ and $\xi^j_{DL,y}$ than the Pt 8/FeCoB 8 and the Ta 5/FeCoB 8, but a factor of 2 smaller resistivity than the Ta device (200 μΩ cm). The combination of the high $\xi^j_{DL,z}$, $\xi^j_{DL,y}$, and low resistivity of the PtTi-based z-spin device should benefit the current and energy efficiencies, impedance, operation speed, and endurance of the SOT devices in memory and logic technologies.

In Fig. 3c, we further show that both $\xi^j_{DL,z}$ and $\xi^j_{DL,y}$ of the PtTi 8/FeCoB 8 do not degrade by annealing up to 400 °C, fulfilling the thermal stability requirement of CMOS integration. In Fig. 3d, we demonstrate complete, deterministic switching of a typical PtTi 5.6/Ti 0.8/FeCoB 1.3 Hall-bar device (with large PMA field of 490 mT, high coercivity $H_c$ of 28 mT, and $M_s$ of 1190 kA/m$^3$, Fig. S3) in the absence of any external magnetic field by an in-plane charge current of ≈ 7.9 mA (or current density of 1.9×10$^7$ A/cm$^2$ within the PtTi layer) injected in a "C-shaped" contact configuration. Here, the 0.8 nm Ti spacer layer is used to enhance the PMA of the FeCoB layer.

However, the switching current density within the spin-current generating layer ($j_c$) is not always a fair indicator for the energy efficiency of practical nanodevices because, despite the width ($W$) of the device can be scaled down, the layer thicknesses may not be scaled for most structures due to the thickness dependences of $j_c$, the SOTs, resistivity, current shunting into the ferromagnets. Thus, we parameterize the energy efficiency by $(I_c/W)^2 \rho_{xx}$ rather than simply $j_c^2 \rho_{xx}$. As compared in Fig. 3e, the PtTi /Ti/FeCoB device exhibits the lowest energy efficiency and the highest switching ratio among the stable perpendicular SOT devices when switched by $z$ spin-current in the absence of any external magnetic field, including Ta/FeCoB ($H_c$ = 20 mT, electric asymmetries),[34] in-plane FeCoB/Ti/ perpendicular FeCoB ($H_c$ = 2 mT, orthogonal trilayer),[24] and CuPt/CoPt ($H_c$ = 24.8 mT, low-symmetry magnetic crystal),[10] Mn$_3$Sn/Cu/[Ni/Co]$_3$ ($H_c$ = 10.8 mT, low-symmetry antiferromagnetic crystal),[31] Mn$_3$Pt/Ti/FeCoB ($H_c$ = 6.1 mT, low-symmetry antiferromagnetic single crystal)[12], RuO$_2$/Ru/Co/Pt ($H_c$ = 40 mT, low-symmetry altermagnetic crystal)[30], Mn$_2$Au/Mo/FeCoB ($H_c$ = 82 mT, low-symmetry antiferromagnetic crystal),[13] and Mn$_3$SnN/[Co/Pd]$_3$ ($H_c$ = 1.4 mT, low-symmetry antiferromagnetic crystal).[27]

The PtTi/Ti/FeCoB device with a high coercivity ($H_c$ =28 mT) is also advantageous over the low-coercivity architectures of FeCoB/Ti/FeCoB (2 mT)[24], Mn$_3$SnN/[Co/Pd]$_3$ (1.4 mT),[27] Mn$_3$Sn/Cu/[Ni/Co]$_3$ (10.8 mT),[31] and Mn$_3$Pt/Ti/FeCoB (6.1 mT)[12]. This is because a perpendicular SOT device must have a sufficiently high coercivity to achieve the required magnetic stability in the environment although it linearly increases the switching current within the switching models of macrospin or domain wall depinning.[1]

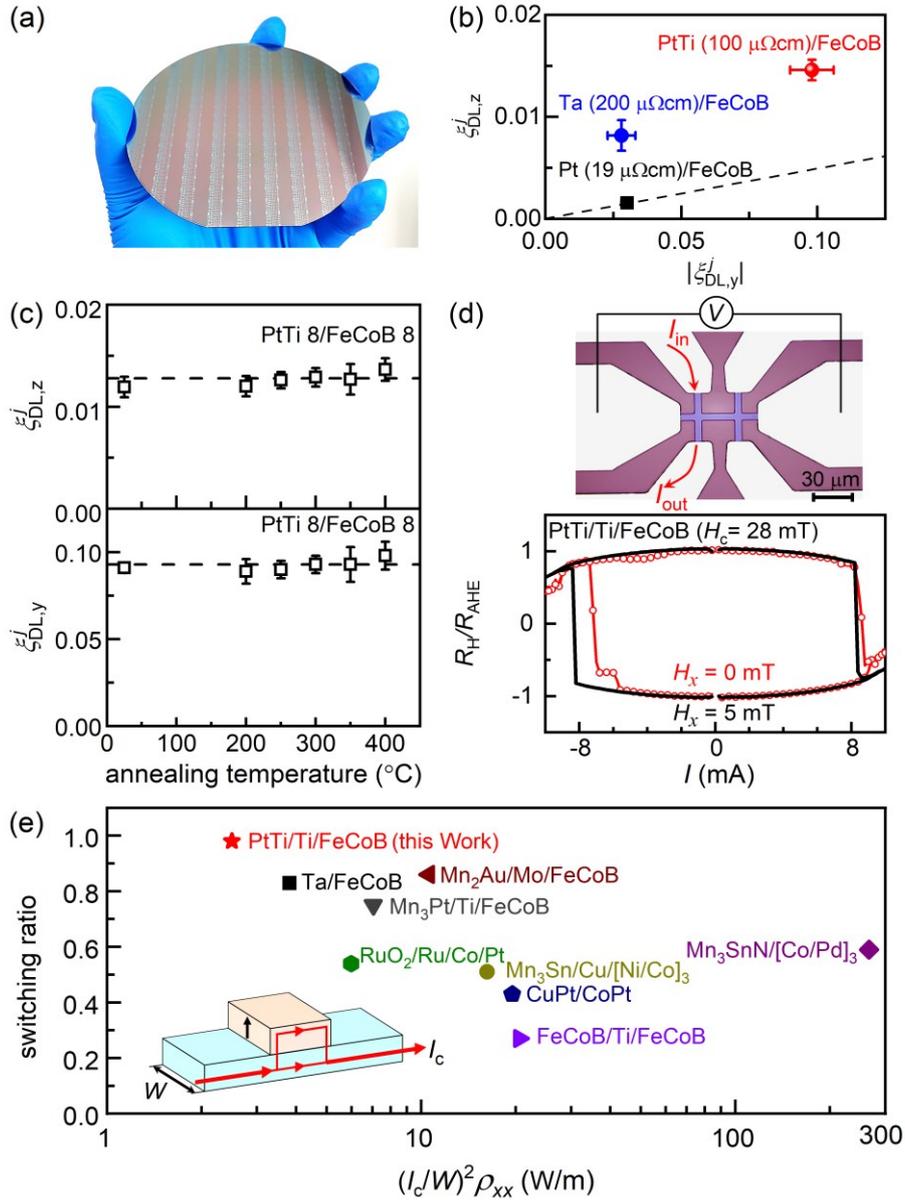

**Fig. 3 Technological importance.** (a) Wafer-scale fabrication (4 inch oxidized silicon wafer) of the z-spin devices of PtTi/Ti/FeCoB heterostructure with strong perpendicular magnetic anisotropy and high coercivity. (b) Comparison of $\xi^j_{DL,z}$, $\xi^j_{DL,y}$, and resistivity of the PtTi 8/FeCoB 8, Pt 8/FeCoB 8, and Ta 4/FeCoB 8.1 devices. (c) $\xi^j_{DL,z}$ and $\xi^j_{DL,y}$ of the PtTi 8/FeCoB 8 as a function of the annealing temperature. In (b) and (c), $W$=10 μm, $L$=20 μm. (d) Optical microscopy image of the Hall-bar device, the C-shaped current injection configuration, and current-driven switching of the PtTi 5.6/Ti 0.8/FeCoB 1.3 with (black) and without external magnetic field (red), suggesting a complete field-free switching. (e) Comparison of the switching ratio and the power parameter of the representative z-spin Hall-bar devices, including the PtTi 5.6/Ti 0.8/FeCoB 1.3, Ta/FeCoB,[34] in-plane FeCoB/ Ti/perpendicular FeCoB[24], CuPt/CoPt[10], Mn$_3$Sn/Cu/ [Ni/Co]$_3$,[31] Mn$_3$Pt/Ti/FeCoB,[12] RuO$_2$/Ru/Co/Pt,[30] Mn$_2$Au/Mo/FeCoB,[13] and Mn$_3$SnN/[Co/Pd]$_3$.[27]

**Conclusion**

We have demonstrated giant enhancement of perpendicular z spin generation by alloying the strongest spin Hall metal Pt with Ti and by electric symmetry engineering, without the need for any thickness or composition gradients. The optimal Pt$_{75}$Ti$_{25}$/FeCoB devices exhibit a factor of 6 and 3 greater dampinglike SOT efficiencies of z spins and y spins, and enable complete deterministic switching of perpendicularly magnetized FeCoB devices at record-low power consumption. The combination of the highly efficient simultaneous generation of z and y spin currents without the need for any thickness or composition gradients, relatively low resistivity, wafer-scale uniform sputter-deposition on silicon oxide, and excellent thermal stability of exceeding 400 ºC make the Pt$_{75}$Ti$_{25}$/FeCoB heterostructure a very compelling candidate for scalable, energy-efficient, field-free, and CMOS-compatible spin-torque technologies. These results establish an effective strategy to develop high-performance z spin generators, especially by alloying and electric asymmetry engineering. We have also clarified that, without altering the spin Hall conductivity tensor, a tilted current flow can never generate any z spin current because none of the three components of the electric field ($\vec{E_x}$, $\vec{E_y}$, $\vec{E_z}$) can generate a spin current both polarized and flows in the z direction.

## Methods

**Sample fabrications:** Samples for this work include spin-torque ferromagnetic resonance (ST-FMR) devices of PtTi 4, 6, 8/FeCoB 2-10, Pt 8/FeCoB 8, and Ta 5/FeCoB 8 with in-plane magnetic anisotropy and a PtTi 5.6/Ti 0.8/FeCoB 1.3 Hall-bar device with perpendicular magnetic anisotropy (PtTi=Pt$_{75}$Ti$_{25}$, FeCoB = Fe$_{60}$Co$_{20}$B$_{20}$, the numbers are layer thicknesses in nanometers). Each device was sputter-deposited on an oxidized silicon substrate and protected by a MgO 1.6/TaO$_x$ 1.6 bilayer. The Pt$_{75}$Ti$_{25}$ layers are grown by co-sputtering using Pt and Ti targets. The base pressure during deposition is below 8×10$^{-9}$ Torr. All the devices are patterned using photolithography and ion milling, followed by the deposition of Ti 5/Pt 150 as the electrical contacts. Annealing experiments were performed in vacuum at different temperatures for 30 min.

**Finite-element analysis**: The electric asymmetries and the relative electric field distributions are simulated by finite-element analysis for a heavy metal/FeCoB bilayer device (with the length $L$, the width $W$, the FeCoB thickness $t$, and the HM resistivity $\rho_{xx}$) covered by two contact pads with a dimension of 20μm×40μm×150nm and a resistivity of 24 μΩ cm. The distribution of charge current density/electric field is calculated following the current continuity equation $\nabla \cdot \boldsymbol{j_c} + \frac{\partial \rho}{\partial t} = 0$ and the boundary condition of $\boldsymbol{n} \cdot \boldsymbol{j_c} = 0$, where $\boldsymbol{j_c} = \boldsymbol{E}/\rho_{xx}$ is the charge current density, $\rho$ is the charge density, $n$ is the normal direction of the device boundary. The resistivity of the FeCoB is fixed at 130 μΩ cm following our resistivity calibration. The element size is 200 nm (length)×200 nm (width)×0.16 nm (thickness).

**ST-FMR analysis:** The magnitudes of $S_{DL,y}$, $S_{DL,z}$, $A_{FL,y}$, and $A_{DL,z}$ in Equations (2) and (3) correlate to the magnitudes of $H_{DL,y}$, $H_{DL,z}$, $H_{FL,y}$, and $H_{FL,z}$ as

$S_{DL,y} = I_{rf} C_{MR} H_{DL,y}$, (M1)

$S_{DL,z} = I_{rf} C_{MR} (\hbar/2e) H_{DL,z}$, (M2)

$A_{FL,y} = I_{rf} C_{MR} (H_{FL,y} + H_{Oe,y})\sqrt{1 + M_{eff}/H_r}$,

$= I_{rf} C_{MR} (\frac{\hbar}{2e} j_{HM} \frac{\xi_{FL,y}}{\mu_0 M_s t_{FM}} + \frac{j_{HM} d_{HM}}{2})\sqrt{1 + M_{eff}/H_r}$, (M3)

$A_{DL,z} = I_{rf} C_{MR} H_D \sqrt{1 + M_{eff}/H_r}$, (M4)

where $C_{MR}$ is a coefficient related to the magnetoresistance of the magnetic layer, and $H_{Oe,y}$ is the Oersted field.

**Acknowledgment:** This work was supported partly by the National Key Research and Development Program of China (2022YFA1204000), the Beijing National Natural Science Foundation (Z230006), and the National Natural Science Foundation of China (12304155, 12274405).

**Data and Materials Availability:** All data needed to evaluate the conclusions in the paper are present in the paper and/or the Supplementary Materials.